\documentclass[superscriptaddress,preprint,prl]{revtex4-1}
\usepackage{graphicx}
\usepackage{amsmath}
\usepackage{amssymb}
\usepackage[squaren, thinqspace]{SIunits}
\usepackage{xspace}
\usepackage{setspace}
\usepackage{color}

\renewcommand{\degree}{\ensuremath{^\circ}\xspace}

\newcommand{\x}{\ensuremath{\mathbf{x}}\xspace}

\newcommand{\Smag}{\ensuremath{\left|S_{21}\right|}\xspace}

\begin{document}

\title{Elastically driven ferromagnetic resonance in nickel thin films}


\author{M. Weiler}
\affiliation{Walther-Mei{\ss}ner-Institut, Bayerische Akademie der Wissenschaften, 85748 Garching, Germany}
\author{L. Dreher}
\affiliation{Walter Schottky Institut, Technische Universit\"{a}t M\"{u}nchen, 85748 Garching, Germany}
\author{C. Heeg}
\author{H. Huebl}
\author{R. Gross}
\affiliation{Walther-Mei{\ss}ner-Institut, Bayerische Akademie der Wissenschaften, 85748 Garching, Germany}
\author{M.S. Brandt}
\affiliation{Walter Schottky Institut, Technische Universit\"{a}t M\"{u}nchen, 85748 Garching, Germany}
\author{S.T.B. Goennenwein}
\email{goennenwein@wmi.badw.de}
\affiliation{Walther-Mei{\ss}ner-Institut, Bayerische Akademie der Wissenschaften, 85748 Garching, Germany}

\begin{abstract}
Surface acoustic waves (SAW) in the GHz frequency range are exploited for the all-elastic excitation and detection of ferromagnetic resonance (FMR) in a ferromagnetic/ferroelectric (Ni/LiNbO$_3$) hybrid device. We measure the SAW magneto-transmission at room temperature as a function of frequency, external magnetic field magnitude, and orientation. Our data are well described by a modified Landau-Lifshitz-Gilbert approach, in which a virtual, strain-induced tickle field drives the magnetization precession. This causes a distinct magnetic field orientation dependence of elastically driven FMR that we observe in both model and experiment.
\end{abstract}

\maketitle

Inverse magnetostriction, or magnetoelasticity, enables the control of the magnetization in ferromagnetic materials via elastic stress~\cite{zheng:2004, spaldin:2005, Eerenstein:2007, Bihler:2008, rushforth:2008, Weiler:2009}. This spin-mechanical interaction prevails at radio frequencies (RF), so that magnonic and phononic degrees of freedom become coupled, as  discussed theoretically~\cite{Tiersten:1964,Kobayashi:1973,Kobayashi:1973-1,Fedders:1974}. We here apply this concept to a surface acoustic wave (SAW) traversing a ferromagnetic thin film. Due to magnetoelastic coupling~\cite{boemmel:1959}, the elastic deformation periodic in time and space results in a change of the magnetic anisotropy, which in turn exerts a torque on the magnetization. Since typical SAW frequencies range from a few MHz to several GHz, SAW-based RF spin-mechanics should enable the study of magnetization dynamics such as ferromagnetic resonance, driven only via RF elastic deformation, not via externally applied RF magnetic fields. The interaction of SAWs and ferromagnetic thin films has been studied experimentally by several groups~\cite{Ganguly:1976,Feng:1982,Wiegert:1987,Wiegert:2001}.  Based on SAW magneto-transmission measurements with the static external magnetic field applied either perpendicular or parallel to the SAW propagation direction, these authors suggested that a magnetoelastic interaction most probably was the dominant interaction mechanism. However, conclusive evidence for the occurrence of an elastically driven, acoustic FMR has proven elusive, and important aspects of the interaction mechanism still await explanation, as stated by Wiegert as recently as 2002~\cite{Wiegert:2002}.  Our experimental study of SAW-based RF spin mechanics as a function of magnetic field magnitude and orientation provides clear evidence for elastically excited, acoustic ferromagnetic resonance. Our findings thus extend the application and understanding of magnetoelastic interaction phenomena in the RF regime.

\begin{figure}
  \includegraphics{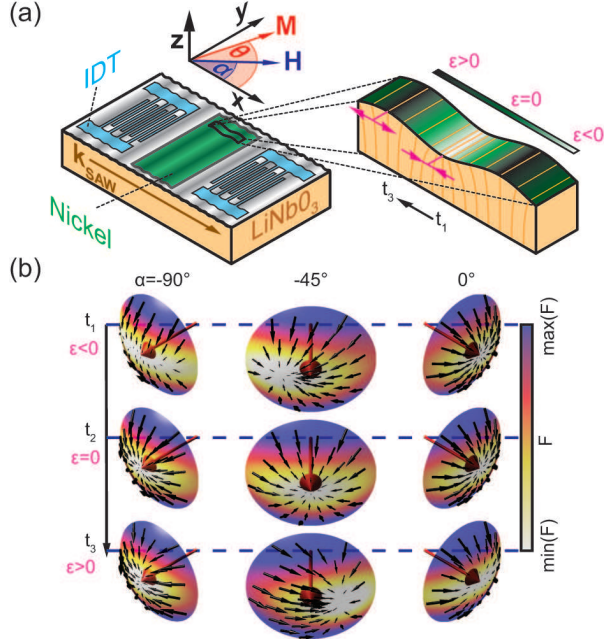}\\
  \caption{(color online) Principles of elastically driven ferromagnetic resonance. (a) Illustration of experimental setup and coordinate system. The closeup to the right shows the strain $\varepsilon$ in the Nickel thin film. (b) The color coded caps show calculations of the magnetic free energy density $F$. The black arrows depict the gradient of $F$ with respect to the magnetization direction  $\mathbf{m}$. A finite gradient at the equilibrium orientation of $\mathbf{M}$ (red arrow) and thus FMR is found only for $\alpha=-45\degree$ and $\varepsilon \neq 0$.}\label{fig:didaktik}
\end{figure}

We study the hybrid device depicted schematically in Fig.~\ref{fig:didaktik}(a). It consists of a \unit{50}{\nano\meter} thick polycrystalline ferromagnetic nickel film deposited onto the central part of a \unit{5\times6}{\milli\meter\squared} y-cut lithium niobate substrate. \unit{70}{\nano\meter} thick Aluminum interdigital transducers (IDTs)~\cite{datta:SAW} with a finger width of \unit{5}{\micro\meter} are employed to launch and detect a SAW. The complex forward transmission $S_{21}$ of the approximately \unit{850}{\micro\meter} long delay line is determined using vector network analysis with a time domain window to cancel the contributions of electromagnetic crosstalk and triple transit interference from the signal. For the description of the experimental and theoretical results we employ the cartesian coordinate system shown in Fig.~\ref{fig:didaktik}(a), where $\alpha$ is the angle between an external magnetic field $\mathbf{H}$ applied in the Ni film plane and the SAW propagation direction $\mathbf{k}_\mathrm{SAW} \| \x$. We first discuss the influence of the SAW on the magnetization vector $\mathbf{M}$ based on magnetoelastic coupling. In the static limit, the magnetic free energy density $F^{0}$ of the Ni film normalized to the saturation magnetization $M_{\mathrm{s}}$  is given by~\cite{chikazumi:ferromagnetism}
\begin{equation}\label{eq:FDC}
    F^{0}(\mathbf{m})=-\mu_0 \mathbf{H} \cdot \mathbf{m}+B_{\mathrm{d}} m_{\mathrm{z}}^2+B_{\mathrm{u}}\left(\mathbf{u}\cdot\mathbf{m}\right)^2 +const.\;,
\end{equation}
where $\mathbf{m}=\mathbf{M}/M_\mathrm{s}$ with components \{$m_\mathrm{x}, m_\mathrm{y}, m_\mathrm{z}$\}. $B_\mathrm{d}=\mu_0M_{\mathrm{s}}/2$ is the shape anisotropy and $B_{\mathrm{u}}$ is a uniaxial in-plane anisotropy field along $\mathbf{u}=\{u_\mathrm{x}, u_\mathrm{y},0\}$. In equilibrium, the magnetization is oriented along a minimum of  $F^{0}$. Due to the shape anisotropy, and as we only consider $\mathbf{H}$ in the film plane here, the equilibrium $\mathbf{m}$ orientation is in the film plane, at an angle $\theta_0$ between \x and $\mathbf{m}$.

The SAW generates an RF strain $\varepsilon$ in the ferromagnetic thin film plane. Due to magnetoelasticity~\cite{chikazumi:ferromagnetism}, this strain results in an RF contribution $F^{\mathrm{RF}}$ to the magnetic free energy density
\begin{equation}\label{eq:FAC}
    F^{\mathrm{RF}}(\mathbf{m})=B_{1}\varepsilon\left(\mathbf{x},t\right) m_\mathrm{x}^2+const.\;,
\end{equation}
where $B_1$ is the magnetoelastic coupling constant and shear strains have been neglected for simplicity. The effective field $\mathbf{H}_\mathrm{eff}$ acting on $\mathbf{m}$ is given by $-\nabla_\mathbf{m} F= -\nabla_\mathbf{m} \left( F^{0}+ F^{\mathrm{RF}} \right)$ ~\cite{Fidler:2000,Pechan:2001}, evaluated at the equilibrium orientation $\theta_0$.

The qualitative effect of the SAW on the magnetization is illustrated in Fig.~\ref{fig:didaktik}(b), where we exemplarily consider an in-plane magnetically isotropic nickel film ($B_{\mathrm{u}}=0$). The total magnetic free energy density $F$ is depicted by the color code in the caps  close to the magnetization equilibrium position $\theta_0=\alpha$ for $\alpha \in \left\{-90\degree,-45\degree,0\degree\right\}$ and $\varepsilon<0$, $\varepsilon=0$ and $\varepsilon>0$. The "tickle field", i.e., the components of $\mathbf{H}_\mathrm{eff}$ perpendicular to $\mathbf{M}$, is depicted by the black arrows on top of the caps. One observes from Fig.~\ref{fig:didaktik}(b) that the tickle field strongly depends on the external static magnetic field orientation $\alpha$ and vanishes for $\alpha=0\degree°$ and $\alpha=90\degree°$. Thus, for $B_{\mathrm{u}}=0$, the acoustic FMR signal will vanish for $\alpha=n\cdot90\degree, n\in\mathbb{Z}$, resulting in a four-fold symmetry. This is in stark contrast to conventional FMR, in which the RF magnetic field is applied externally and does \emph{not} depend on $\alpha$. Taken together, the above model suggests that the characteristic dependence on the orientation of the external static magnetic field can be used as a fingerprint to distinguish acoustically driven FMR from conventional FMR.

\begin{figure}
  \includegraphics{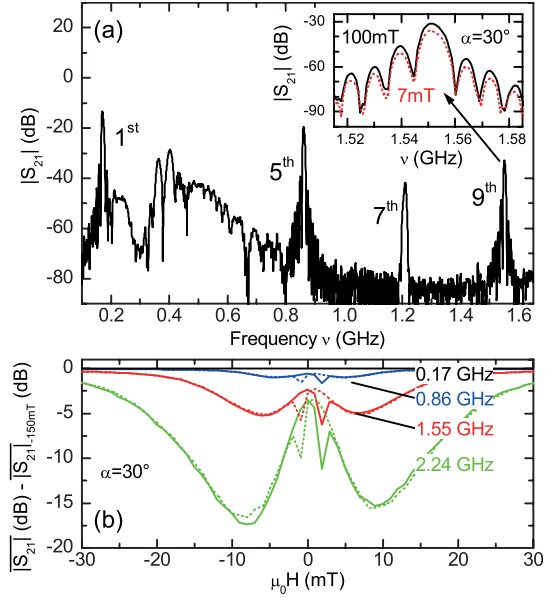}\\
  \caption{(color online) Experimental observation of elastically driven FMR in SAW transmission. (a) \Smag as a function of frequency at $H=0$. The inset shows \Smag around the 9$^{\mathrm{th}}$ harmonic, recorded with the external magnetic field applied at $\alpha=30\degree$, for $\mu_0|H|=100\,\mathrm{mT}$ (solid line) and $\mu_0|H|=7\,\mathrm{mT}$ (dotted line). (b) $\overline{\left|S_{21}\right|}$ as a function of the external magnetic field applied at $\alpha=30\degree$ (solid lines: upsweep, dashed lines: downsweep). With increasing frequency, the damping of the SAW increases and the field-symmetric absorption dips shift to higher magnetic fields, as expected for FMR.}\label{fig:freq}
\end{figure}
We next discuss our experimental results. Figure~\ref{fig:freq}(a) shows the magnitude of the transmission \Smag measured in the nickel/lithium niobate hybrid as a function of frequency at zero external magnetic field. Several RF transmission maxima, corresponding to propagating SAWs, are observed. The first transmission maximum occurs at the SAW delay line fundamental frequency of \unit{172}{\mega\hertz}. SAW transmission is also observed at odd harmonics, which allows choosing the magnetoelastic interaction frequency in a range from \unit{172}{\mega\hertz} to \unit{3.6}{\giga\hertz}. Here, we constrain our discussion to the fundamental frequency as well as the 5$^{\mathrm{th}}$, 9$^{\mathrm{th}}$ and 13$^{\mathrm{th}}$ harmonic frequency, as these frequencies exhibit the most intense SAW transmission for the chosen IDT metallization ratio of 0.5. Using appropriate IDT designs~\cite{datta:SAW} it is possible to excite SAWs at several almost arbitrarily chosen frequencies, or even a broad frequency range. The inset in Fig.~\ref{fig:freq}(a) exemplarily shows the frequency dependence of \Smag around the 9$^{\mathrm{th}}$ harmonic (\unit{1.55}{\giga\hertz}) for two different values of the external magnetic field. The series of transmission lobes is characteristic of a SAW delay line passband~\cite{datta:SAW}. As $\mu_0|H|$  is changed from \unit{100}{\milli\tesla} to \unit{7}{\milli\tesla} the damping of the SAW increases by approximately \unit{5}{dB}. Figure~\ref{fig:freq}(b) depicts the evolution of the SAW transmission as a function of magnetic field strength for a full magnetic field sweep from \unit{-150}{\milli\tesla} to \unit{+150}{\milli\tesla} and back to \unit{-150}{\milli\tesla}. The data correspond to the SAW transmission averaged over the FWHM of the central passband lobe $\overline{\left|S_{21}\right|}$. In addition to the hysteretic magnetization switching at $\left|\mu_0H\right|<\unit{3}{\milli\tesla}$ already reported~\cite{Feng:1982,Wiegert:1987}, two absorption maxima can be discerned as pronounced dips in Fig.~\ref{fig:freq}(b). The latter show no hysteresis, are symmetric to zero external magnetic field, and are both present in the magnetic field up- as well as the downsweep. Furthermore, the maximal SAW attenuation increases with frequency from less than \unit{0.1}{dB} at \unit{0.17}{\giga\hertz} to approximately \unit{16}{dB} at \unit{2.24}{\giga\hertz}, and simultaneously shifts to larger $H$. Due to the characteristic non-hysteretic behavior of the attenuation maxima together with their shift as a function of frequency, we attribute the attenuation maxima to FMR. The slight asymmetry in the attenuation maxima in Fig.~\ref{fig:freq}(b) is observed for all $\mathbf{H}$ orientations. Further work will be necessary to determine its origin.

\begin{figure}
  \includegraphics{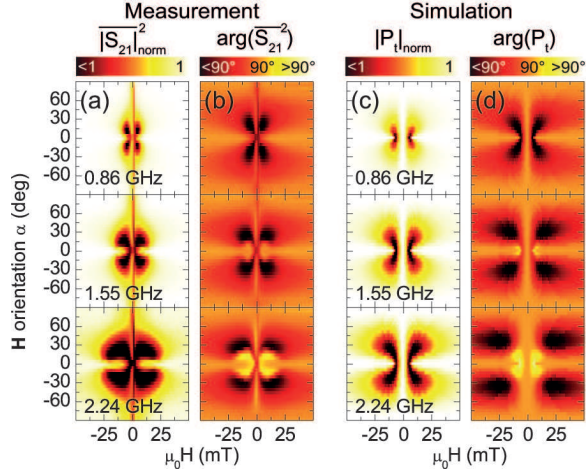}\\
  \caption{(color online) Evolution of the SAW transmission as a function of the magnitude and orientation of the external magnetic field. Column (a) shows the experimentally determined $\overline{\left|S_{21}\right|}^2_\mathrm{norm}$. In resonance, the SAW transmission is attenuated, as evidenced by the black color. Column (b) shows the experimentally determined $\arg (\overline{S_{21}}^2)$. Column (c) and column (d) show the transmitted SAW power density, simulated using Eq.~\eqref{eq:power}. The simulation reproduces all the salient features observed in the experiment and demonstrates that the FMR is elastically driven.}\label{fig:contour}
\end{figure}

To identify the mechanism driving the FMR, a detailed study of the angular dependence is mandatory, as discussed in the context of Fig.~\ref{fig:didaktik}(b). We thus performed a series of measurements to determine $\overline{\left|S_{21}\right|}$ as a function of magnetic field magnitude and orientation. The orientation $\alpha$ of the in-plane static magnetic field was varied in $5\degree$ steps from $-90\degree$ to $90\degree$. Figure~\ref{fig:contour}(a) shows the $\overline{\left|S_{21}\right|}^2$ data obtained as a false color plot. The color code represents the normalized transmission $\overline{\left|S_{21}\right|}^2_\mathrm{norm}=\overline{\left|S_{21}\right|}^2(\mu_{0}H) / \overline{\left|S_{21}\right|}^2(-150\,\mathrm{mT})$, with black indicating maximal absorption. Figure ~\ref{fig:contour}(b) depicts the corresponding signal phase $\arg (\overline{S_{21}}^2)$, setting $\arg (\overline{S_{21}}^2)(\mu_{0}H=\unit{-150}{\milli\tesla})=90\degree$, with black and yellow visualizing deviations of the signal phase from 90\degree (orange). As evident from Fig.~\ref{fig:contour}, the SAW transmission exhibits a characteristic, approximately four-fold angular symmetry (four transmission minima as a function of $0\degree\le\alpha\le360\degree$ for a given external magnetic field magnitude). Thus, in addition to the shift of the resonance field with frequency already discussed in the context of Fig.~\ref{fig:freq}, the anisotropy of the SAW magneto-transmission exhibits the characteristic fingerprint of SAW-driven acoustic FMR. In particular, no attenuation occurs for $\alpha=0\degree$ and $\alpha=\pm90\degree$, as expected according to the picture shown in Fig.~\ref{fig:didaktik}(b). The fact that the attenuation maxima are not located exactly at $\pm 45\degree$ is due to a uniaxial in-plane magnetic anisotropy $B_\mathrm{u} \neq 0$, as discussed in the following.

To model the SAW magneto-transmission, we use a Landau-Lifshitz-Gilbert (LLG) approach, taking into account RF magnetoelastic effects. We neglect spatial variations of the strains and restrict the calculations to the collective FMR mode. The LLG equation~\cite{Landau:1935,Gilbert:2004}
\begin{equation}\label{eq:LLG}
    \partial_t \mathbf{m}=\gamma \mathbf{m} \times \mu_0 \mathbf{H}_\mathrm{eff} + a \mathbf{m} \times \partial_t \mathbf{m}\;,
\end{equation}
describes the evolution of the magnetization direction $\mathbf{m}$ in an effective magnetic field $\mathbf{H}_\mathrm{eff}$, with $\gamma$ and $a$ being the gyromagnetic ratio and the Gilbert damping parameter, respectively. We solve the LLG in a cartesian coordinate system with the first axis pointing along the equilibrium direction of $\mathbf{m}$, which in our case is in the film plane at an angle $\theta_0$ to the \x axis and the third axis pointing out-of-plane. In the following, the subscripts \{1,2,3\} refer to this frame of reference. Thus, $\mathbf{m}=\left\{1,m_2,m_3\right\}$, where $m_2$ and $m_3$ denote small deviations from equilibrium. The non-vanishing component of the effective RF magnetic field $\mathbf{h}(t)=-\nabla_\mathbf{m} F^\mathrm{RF}(t)|_{\theta=\theta_0}$~\cite{Fidler:2000,Pechan:2001} is given by
\begin{equation}\label{eq:hAC}
    \mu_0 h_2(t)=-2B_1 \varepsilon(t) \cos(\theta_0) \sin(\theta_0)\;.
\end{equation}
In contrast to classical FMR, where an external RF magnetic field is exploited to drive the resonance, Eq.~\ref{eq:hAC} shows that acoustic FMR is excited by a purely internal RF magnetic field $\mathbf{h}(t)$. The latter is due to RF spin mechanics, i.e., it is generated by the magnetoelastic interaction of the SAW elastic strain field with the ferromagnet. This coupling between $\mathbf{m}$ and the SAW results in a resonant attenuation and phase shift of the SAW transmission when the condition for ferromagnetic resonance is met. We write the transmitted RF power density $P_\mathrm{t}$ as
\begin{equation}\label{eq:power}
   P_\mathrm{t} \propto P_0 - \frac{1}{2}\mu_0\omega\mathbf{h}^\mathrm{\dagger} \stackrel{\leftrightarrow}{\chi} \mathbf{h}\;,
\end{equation}
where the Polder susceptibility $\stackrel{\leftrightarrow}{\chi}$ is a function of the static magnetic free energy density $F^{0}$, $P_0$ is the power density transmitted off resonance and $\omega=2\pi \nu$ is the SAW angular frequency.  The non-vanishing components of the Polder susceptibility tensor read as $\chi_{22}=M_s\gamma\mu_0\omega_1/D$, $\chi_{33}=M_s\gamma\mu_0\omega_2/D$ and $\chi_{23}=\chi_{32}^*=M_s\gamma\mu_0i\omega/D$ with $D=\omega_{1}\omega_{2}-\omega^2$, $\omega_{1}=\gamma\left( \mu_{0}H_1 + 2B_{\mathrm{d}} - 2 B_{\mathrm{u}} u_{1}^{2} \right) + i\omega a$ and $\omega_{2}=\gamma\left(\mu_{0}H_1 + 2 B_{\mathrm{u}}\left( u_{2}^{2} - u_{1}^{2}\right)\right) + i\omega a$. Eq.~\eqref{eq:power} corresponds to the conventional FMR formula~\cite{Kobayashi:1973}, but employs the purely virtual $\mathbf{h}(t)$ defined in Eq.~\eqref{eq:hAC} instead of an externally applied, real RF magnetic field.

Figure~\ref{fig:contour}(c) shows $|P_{\mathrm{t}}|_\mathrm{norm}=|P_{\mathrm{t}}|(\mu_0H)/|P_{\mathrm{t}}|(\unit{-150}{\milli\tesla})$ calculated using Eq.~(\ref{eq:power}), with $B_\mathrm{d}=\unit{+300}{\milli\tesla}$, $a=0.25$, $\varepsilon=10^{-5}$ and $B_\mathrm{u}=\unit{+4}{\milli\tesla}$ along the \x-direction for the three frequencies investigated in the experiment. Figure~\ref{fig:contour}(d) depicts the corresponding transmission phases $\arg (P_\mathrm{t})$. The simulation reproduces all the salient features observed in the experiment, in particular the angular dependence. Considering our rather simple model, the agreement is excellent.  This demonstrates that the SAW absorption is indeed caused by acoustically driven ferromagnetic resonance, i.e., by RF magnetoelastic interactions, and not by conventional FMR. Moreover, it corroborates the use of a virtual, purely internal $\mathbf{h}(t)$ on the same footing as a real externally applied magnetic field in Eq.~\eqref{eq:power} a posteriori. We note that our damping constant $a=0.25$ is about a factor of ten larger than that expected from cavity FMR experiments~\cite{Bhagat:1966}. This value was chosen to account for line broadening presumably due to the inhomogeneous tickle field periodic with the SAW wavelength $\lambda\approx\unit{1.5}{\micro\meter}$ at \unit{2.24}{\giga\hertz}, corresponding to a wavevector $k_\mathrm{SAW}\approx\unit{4\times10^6}{\meter^{-1}}$. In coplanar waveguide FMR such non-uniform excitation fields are known to lead to a line broadening by 100\% already for $k=\unit{6\times10^4}{\meter^{-1}}$~\cite{Counil:2004}. The much larger wavevectors in our case will therefore lead to considerably higher linewidths, i.e. $a=0.25$. To circumvent such inhomogeneous broadening effects, one may miniaturize the ferromagnetic thin film to lateral dimensions much smaller than the SAW wavelength. In such samples, a determination of the influence of magnon-phonon interactions on magnetic damping should be possible.

In conclusion, we have consistently found both in experiment and in simulation that the magnetoelastic interaction of a SAW with a ferromagnetic thin film allows to excite FMR in the film. The FMR  is driven acoustically in the sense that no external RF magnetic field is applied to the ferromagnet. Rather, a purely internal RF magnetic field arises due to magnetoelastic coupling between the SAW elastic strain field and the ferromagnet. Using a free energy approach as well as LLG calculations, we showed that the magnitude of the "acoustic" RF magnetic field characteristically depends on the orientation of the magnetization. The angular dependence of the SAW transmission thus is a fingerprint of acoustic FMR, as observed in our experiments. Our experimental findings open a third alternative for the excitation of FMR, in addition to externally applied RF magnetic fields~\cite{Kittel:1948,Bickford:1950} or the spin torque effect~\cite{Slonczewski:1996,Myers:1999,Katine:2000,Stiles:2002,Zhang:2002}. More fundamentally, acoustic FMR can provide a pathway for the study of magnon-phonon interaction phenomena studied predominantly on theoretical grounds to date~\cite{Tiersten:1964,Kobayashi:1973,Kobayashi:1973-1,Fedders:1974}, or also for mechanical spin pumping~\cite{kovalev:2005, mosendz:2010}. The coupling between elastic and magnetic degrees of freedom will open additional channels for spin dephasing, so that the magnetization damping is linked to the magnon-phonon coupling strength. Furthermore, a detailed study of the interaction between magnonic and phononic dispersions, of the strength of the coupling, of the anticrossing of these branches and of the generation of strongly coupled elastic/magnetic states are appealing challenges for the future.

We gratefully acknowledge stimulating discussions with B. Botters, R. Huber and D. Grundler. This work is financially supported by the DFG via project nos GO 944/3-1, SFB 631 C3, and by the Cluster of Excellence Nanosystems Initiative Munich (NIM).

\end{document}